\documentclass[english,aps, prl, 10pt, twocolumn, longbibliography, showpacs,superscriptaddress]{revtex4-1}
\usepackage[T1]{fontenc}
\usepackage[latin9]{inputenc}
\usepackage{graphicx}

\makeatletter
\usepackage[colorlinks=true,bookmarks=false,linkcolor=black,urlcolor=black,citecolor=black,breaklinks]{hyperref}
\usepackage{amsmath}
\usepackage{amsfonts}
\usepackage{graphicx}
\usepackage{braket}
\usepackage[note-name=, use-sort-key =false]{notes2bib}

\makeatother

\usepackage{babel}
\begin{document}

\title{Photon-blockade induced photon anti-bunching in photosynthetic Antennas
with cyclic structures}

\author{Hui Dong}

\affiliation{The Institute for Quantum Science and Engineering, Texas A\&M University,
College Station, Texas}

\author{Sheng-Wen Li}

\affiliation{The Institute for Quantum Science and Engineering, Texas A\&M University,
College Station, Texas}

\author{Zhenhuan Yi}

\affiliation{The Institute for Quantum Science and Engineering, Texas A\&M University,
College Station, Texas}

\author{Girish S. Agarwal}

\affiliation{The Institute for Quantum Science and Engineering, Texas A\&M University,
College Station, Texas}

\affiliation{Department of Biological and Agricultural Engineering,Texas A\&M
University, College Station, Texas}

\author{Marlan O. Scully}

\affiliation{The Institute for Quantum Science and Engineering, Texas A\&M University,
College Station, Texas}

\affiliation{Princeton University, Princeton, New Jersey }

\affiliation{Baylor University, Waco, Texas}
\begin{abstract}
One of the important nonclassical effects in quantum optics is the
anti-bunching, which has been observed in a large class of physical
systems - including light-harvesting antennas with cyclic structures.
The units of the ring couple with adjacent ones through dipole-dipole
interactions. We show how this strong dipole-dipole interaction leads
to photon-blockade resulting in the suppression of double excitation
pathway and anti-bunching in photosynthesis systems. The robustness
of the photon blockade is demonstrated against the disorder in the
ring structures. We hypothesis that the effect may be utilized by
light-harvesting systems to avoid damages from excess energy.
\end{abstract}


\maketitle
\narrowtext

The solar radiation is an abundant, clean, and sustainable source
for the our energy needs in conceivable future \cite{Lewis2006Proc.Natl.Acad.Sci.U.S.A.15729--15735,Blankenship2011Science805--809}.
Promising methods of harvesting sunlight are to utilize the biological
light-harvesting machinery \cite{Lewis2006Proc.Natl.Acad.Sci.U.S.A.15729--15735,Zhu2010Annu.Rev.PlantBiol.235--261,Blankenship2011Science805--809},
or combined technology \cite{Liu2015NanoLett.3634--3639,Nichols2015Proc.Natl.Acad.Sci.U.S.A.11461--11466}
of bio-organic and nanomaterial design. To pursue these new methods,
it requires to understand the underlying design principles of both
the high quantum efficiency \cite{Scholes2011Nat.Chem.763--774,Lambert2013,Pleniobook}
as well as the protection mechanism against damage by excess energy
\cite{Mueller2001Plantphysiology1558--1566} in the natural light-harvesting
complexes in photosynthetic system. Many efforts have been put together
recently to unveil these design principle from optical properties
of individual pigment-complexes \cite{Engel2007Nature782--786,Lee2007Science1462--1465,Collini2009science369--373,Aggarwal2013Nat.Chem.964--970,Wientjes2014Nat.Commun.},
and the structural organizations among complexes \cite{Hu1998Proc.Natl.Acad.Sci.U.S.A.5935--5941,Raszewski2008J.Am.Chem.Soc.4431--4446,Caffarri2011Biophys.J.2094--2103,Cleary2013Proc.Natl.Acad.Sci.U.S.A.8537--8542,Bennett2013J.Am.Chem.Soc.9164--9173,Chen2013PRE,Mourokh2015PRE}.
One prominent discovery is symmetric structures in the light-harvesting
complexes \cite{VanAmerongen2000}. For example, the light harvesting
complexes II (LHCII \cite{Caffarri2011Biophys.J.2094--2103}) of the
green plant antenna, has a $C_{3}$ symmetry in the photosystem II
(PSII), while light-harvesting antennas (LHI and LHII \cite{McDermott1995Nature517--521,Koepke1996Structure581--597,Cogdell2006Q.Rev.Biophys.227--324,Yang2010JournalofChemicalPhysics234501})
in the purple bacteria exhibit a $C_{8}$ symmetry, despite small
disorders.

On the other hand, anti-bunching of photon statistic is an iconic
quantum optical phenomena in atomic ensembles, cavity-atom, and artificial
structures\cite{Tian1992PhysicalReviewA,Rebic1999J.Opt.B490,Birnbaum2005Nature,Saffman2010Rev.Mod.Phys.2313,Jau2016Nat.Phys.71--74,Labuhn2016Nature,Gillet2010Phys.Rev.A013837,PhysRevLett.116.243001}.
In natural light-harvesting systems, the anti-bunching was reported
\cite{Wrachtrup2002AIPConferenceProceedings633_470} in the resonance
fluorescence from LHI. It was explained as a collective behavior \cite{Wrachtrup2002AIPConferenceProceedings633_470}
of the whole ring. The single molecular fluorescence experiment \cite{Aggarwal2013Nat.Chem.964--970}
on the artificial synthetic $\pi$-conjugated spoked-wheel also reported
a similar effect. Benefiting from the recent progress of surface-enhanced
spectroscopy, van Hulst et.al has successfully measured the resonance
fluorescence from a single LHII \cite{Wientjes2014Nat.Commun.}. An
anti-bunching was reported in the second-order correlation of resonance
fluorescence. A mechanism of exciton-exciton annihilation (EEA) \cite{Trinkunas2001Phys.Rev.Lett.86_4167}
is proposed to account for the dip of fluorescence at zero time delay.
However, quantum optics studies\cite{Agarwal1977Phys.Rev.A1613,Birnbaum2005Nature}
suggest that other possibilities cannot be ruled out. In this letter,
we propose photon-blockade mechanism to account for the observed phenomena
instead of the two-excitation annihilation originally suggested in
Ref. \cite{Wientjes2014Nat.Commun.}. We also suggest that the two
mechanisms may be distinguished through a temperature-dependent experiment
\cite{Brueggemann2009Chem.Phys.357_140}.

\begin{figure}
\includegraphics[width=8cm]{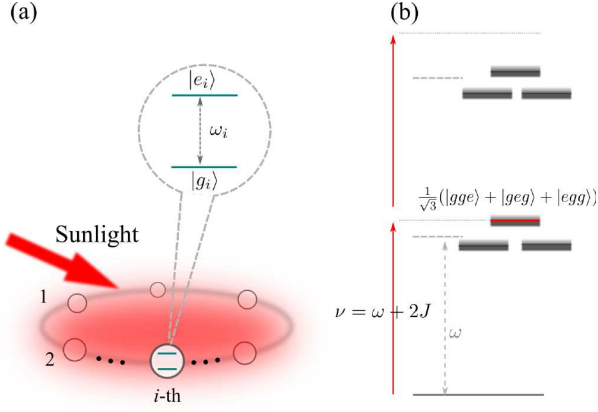}

\caption{( Color online ) (a) Generic ring with $N$ chlorophylls. (b) Example
of energy-levels with $N=3$. The dashed lines show the energy levels,
equally spaced with $\omega$, for a ring without inter-chlorophylls
interaction. The solid lines with shadow represents the energy levels
for the same ring with dipole-dipole interaction.}

\label{fig:setup}
\end{figure}

We consider a generic ring structure ($N$ pigments), wherein the
interaction between chlorophylls is characterized by hopping of excitation
with strength $J$. The configuration is shown in Fig. \ref{fig:setup}
(a), where $N$ chlorophylls are organized evenly on a ring. And the
energy levels ( $\left|e_{i}\right\rangle $ and $\left|g_{i}\right\rangle $)
are illustrated for the $i$th-chlorophyll in the subset of Fig. \ref{fig:setup}(a).
The Hamiltonian reads
\begin{equation}
H_{0}=\sum_{i=1}^{N}\omega_{i}s_{i}^{z}+\sum_{i=1}^{N}J_{i}\left(s_{i}^{+}s_{i+1}^{-}+s_{i+1}^{+}s_{i}^{-}\right),
\end{equation}
where $s_{i}^{+}=\left|e\right\rangle _{i}\left\langle g\right|$,
$s_{i}^{-}=\left|g\right\rangle _{i}\left\langle e\right|$, and $s_{i}^{z}=\frac{1}{2}\left(\left|e\right\rangle _{i}\left\langle e\right|-\left|g\right\rangle _{i}\left\langle g\right|\right)$
with $\left|e\right\rangle _{i}$ ($\left|g\right\rangle _{i}$ )
as the two electronic levels on optical resonance with a laser drive
field or sunlight. The hopping of excitation is induced by the dipole-dipole
interaction \cite{Spano1989J.Chem.Phys.} between chlorophylls, namely
$J_{i}=\left\langle e_{i+1}g_{i}\right|V\left|g_{i+1}e_{i}\right\rangle $,
with 
\begin{equation}
V=\frac{\mu_{i+1}\cdot\mu_{i}-3(\mu_{i+1}\cdot\vec{e}_{R_{i,i+1}})(\mu_{i}\cdot\vec{e}_{R_{i,i+1}})}{|\vec{R}_{i,i+1}|^{3}},
\end{equation}
where $\vec{R}_{i,i+1}$is relatively position between $i$-th and
$\left(i+1\right)$-th chlorophylls and $\vec{e}_{R_{i,i+1}}=\vec{R}_{i,i+1}/|\vec{R}_{i,i+1}|$.
For the interaction, we consider the fact that the distance between
chlorophylls is far smaller the absorption wavelength namely, $|\vec{R}_{i,i+1}|\ll\lambda$,
where $\lambda$ is the absorption wavelength\cite{Spano1989J.Chem.Phys.}.
Typically, the radius of the ring is about 4-5nm for light-harvesting
complexes \cite{Wientjes2014Nat.Commun.} and 6nm for the synthetic
macro-molecule \cite{Aggarwal2013Nat.Chem.964--970}, which is far
smaller than the corresponding absorption wavelength, 800nm \cite{Wientjes2014Nat.Commun.}
and 465nm \cite{Aggarwal2013Nat.Chem.964--970}. We consider in current
model only the nearest neighbor interaction between chlorophylls \cite{Cheng2006Phys.Rev.Lett.028103}. 

For a uniform ring ($\omega_{i}$=$\omega$), the non-perturbative
Hamiltonian $H_{0}$ can be diagnolized via Jordan-Wigner transformation
\cite{Sachdev2011}, where the eigen-system are divided into two categories:
odd and even excitation. The eigen-energies of single and double excitation
states, relative to the ground state with all molecules on ground
states, are given by 
\begin{equation}
E_{\mathrm{s}}\left(k\right)=\omega+2J\cos2\pi k/N,\label{eq:single}
\end{equation}
 and 
\begin{eqnarray}
E_{\mathrm{d}}\left(k_{1},k_{2}\right) & = & 2\omega+2J[\cos2\pi(k_{1}+1/2)/N\nonumber \\
 &  & \qquad\quad\;+\cos2\pi(k_{2}+1/2)/N],\label{eq:double}
\end{eqnarray}
where $k$, $k_{1}$ and $k_{2}\in\{0,1,2,...,N-1\}$, and $k_{1}\neq k_{2}$.
The energy spectrum for a simple case $N=3$ is shown in Fig. \ref{fig:setup}(b).
The energy levels of both single-excitation in Eq. (\ref{eq:single})
and double-excitation in Eq. (\ref{eq:double}) are shifted due to
dipole-dipole interaction between chlorophylls. In current model,
we have assumed all transition dipoles are parallelly pointing up.
This assumption results in a positive coupling strength $J>0$. For
real case of LHI's B800 ring with dipole moments slightly slated,
the coupling is negative. However, the following discussion of off-resonance
is still valid. For the case $J>0$, the highest energy level ($k=0$)
of single excitation subspace is $\left|\psi_{s}\right\rangle =1/\sqrt{N}\sum_{i=1}\left|g_{1}g_{2}...e_{i}...g_{N}\right\rangle $,
which is a single-photon superradiance state. One important feature
is that the amount of energy shift of the superadiance state is larger
than that of highest double-excitation state, as shown in Fig. 1(b).
The transition dipole moment from the ground state to this symmetric
state is enhanced by a factor of $\sqrt{N}$, while the transition
dipoles for other states essentially vanish. Such vanishing of transition
dipole moment leads to the subradiance states \cite{Dicke1954Phys.Rev.99,Scully2015Phys.Rev.Lett.243602},
whose radiation decay rate is significant suppressed. The direct result
of the symmetric structure is the enhancement of transition amplitude
from the ground state to the symmetric state, while transition to
other states are canceled or suppressed. This feature is also known
as super-absorption \cite{Spano1989J.Chem.Phys.,Yang2010JournalofChemicalPhysics234501,Dong2012Light:Sci.Appl.e2}.

As for absorption, we consider the interaction of driving light field
with ring structures. With this observation, we assume all the chlorophylls
couple uniformly to the incident light, and simplify the interaction
of ring with driving field as 
\begin{equation}
H_{I}=\sum_{i=1}^{N}\Omega_{R}\left(s_{i}^{+}e^{-i\nu t}+s_{i}^{-}e^{i\nu t}\right),\label{eq:light_matt_interc}
\end{equation}
where $\Omega_{R}$ is the Rabi frequency associated with coupling
of photon mode with frequency $\nu$ to the atomic transition $\left|g\right\rangle \rightarrow\left|e\right\rangle $. 

The ring structure also interacts with the surrounding vacuum field,
which results in dissipation of excitation in the system. The dissipation
dynamics is described by the master equation as follows \cite{Lehmberg1970Phys.Rev.A883,Agarwal1970Phys.Rev.A2038,Agarwal1977Phys.Rev.A1613}
\begin{equation}
\frac{\partial\rho}{\partial t}=-i\left[\rho,H\right]-\frac{\gamma}{2}\sum_{ij}\left(s_{i}^{+}s_{j}^{-}\rho+\rho s_{i}^{+}s_{j}^{-}-2s_{j}^{-}\rho s_{i}^{+}\right),\label{eq:mastereq}
\end{equation}
where $\gamma$ is dissipation strength, and $H=H_{0}+H_{I}$ is the
total Hamiltonian. Equation (\ref{eq:mastereq}) was used to show
the collective spontaneous emission, known as the Dicke superradiance
\cite{Agarwal1970Phys.Rev.A2038}. In the current discussion, we neglect
collective the Lamb shift \cite{Roehlsberger2010Science} in the master
equation.

\begin{figure}
\includegraphics[width=8cm]{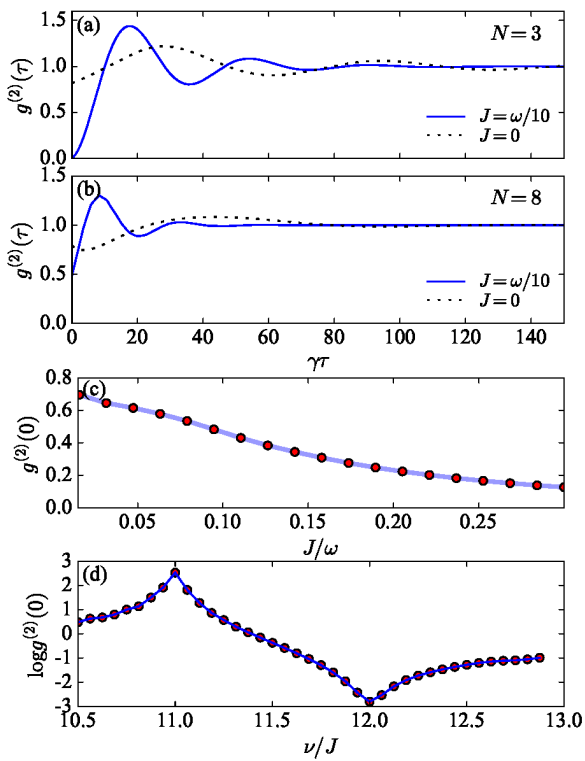}

\caption{(Color online) The second-order correlation functions for system with
(blue solid line) and without dipole-dipole interaction (black dashed
line). The correlations are plotted for the ring structures with (a)
$N=3$ and (b) $8$ chlorophylls. (c) The second-order correlation
at $\tau=0$ is plotted as a function of the strength $J$ of the
dipole-dipole interaction for the ring with 8 chlorophylls. (d) The
second-order correlation at $\tau=0$ is plotted as a function of
the laser frequency $\nu$ for the ring with 3 chlorophylls.}

\label{fig:corr}
\end{figure}

To demonstrate the new mechanism, we mainly focus on calculating the
probability of having double excitations in this ring structure. Experimentally,
the method to determine the number of excitations in a single aggregate
is to measure the multi-order correlations of the resonance fluorescence
\cite{Agarwal1977Phys.Rev.A1613}. To check the probability of double
excitations, we typically measure the second-order correlation. The
measured signal is presented by a normalized second-order correlation
function \cite{Scully1997}, defined as 
\begin{equation}
g^{(2)}(t,\tau)=\frac{\langle S^{+}(t)S^{+}(t+\tau)S^{-}(t+\tau)S^{-}(t)\rangle}{\langle S^{+}(t)S^{-}(t)\rangle\langle S^{+}(t+\tau)S^{-}(t+\tau)\rangle},
\end{equation}
where $S^{\pm}=\sum_{i=1}^{N}s_{i}^{\pm}$. The correlation function
quantifies the probability of having one photon emitted subsequently
at time $t+\tau$, after an initial photon already emitted at time
$t$. A lowing emission probability at $\tau=0$ indicates a suppress
of the probability having double excitation in the system. 

In Fig. \ref{fig:corr}, we show the normalized second-order correlation
$g^{(2)}(\tau)\equiv g^{(2)}(t,\tau)$ as a function of delay time
$\tau$ at steady state ($t\rightarrow\infty$) for the ring structure
with (blue solid curve) and without (black dashed curve) interaction
between adjacent chlorophylls. The correlations are plotted for rings
with 3 and 8 chlorophylls, respectively in subfigures (a) and (b).
The parameters using in the simulation are estimated from LHII in
the purple bacteria \cite{Cheng2006Phys.Rev.Lett.028103,Cleary2013NewJ.Phys.15_125030},
where $J=\omega/10$ and $\gamma=J/10$. We also assume relatively
weak driving field $\Omega_{R}=\gamma$. At $\tau=0$, it shows a
significant reduction of the amplitude of emitting two photons simultaneously,
in comparison with the one without interaction. This phenomena is
known as anti-bunching effect, which can be explained with the energy-level
scheme in Fig. 1(c). Due to the dipole-dipole interaction between
adjacent chlorophylls, the energy for symmetric superadiance state
$\left|\psi_{s}\right\rangle $ is shifted to $\omega+2J$. The interaction
with incident light in Eq. (\ref{eq:light_matt_interc}) can only
induce the transition from the ground state to this superadiance state,
if the transition dipoles of individual chlorophylls are aligned along
the same direction. The probability of exciting the system from superradiance
state to all double-excitation states are suppressed, since a second
photon with the same energy $\omega+2J$ is far detuned from resonance
with the transition from the superradiance state to the double excitation
eigenstates under the condition of strong dipole-dipole interaction.
In Fig. \ref{fig:corr}(c), we show effect of increasing coupling
constant $J$ on the correlation function at zero delay. The curve
shows an enhancement of blockade with a decrease of emission rate\cite{Gillet2010Phys.Rev.A013837},
due to increasing detuning with larger interaction strength. The anti-bunching
depends intensively on how the driving field is tuned. In Fig. \ref{fig:corr}(d),
we show the correlation function $g^{\left(2\right)}\left(\tau=0\right)$
as a function of laser frequency $\nu$ for the ring with $N=3$.
The dip on the curve shows the anti-bunching in resonance fluorescence
($\nu/J=12$), while the peak illustrates photon bunching ( $g^{\left(2\right)}\left(\tau=0\right)>1$
) under the two-photon resonance ($\nu/J=11$) with the transition
from the ground state to double-excitation state. This feature is
unique to the photon-blockade mechanism, and can be utilized to distinguish
the current mechanism from EEA.

The current mechanism is also an analogy to photon-blockade effect
of a single atom in a cavity \cite{Birnbaum2005Nature}, where the
interaction between the cavity mode and the single atom tune all the
double-excitation states off-resonant with the incident photon. Instead
of a single atom \cite{Kimble1977Phys.Rev.Lett.}, we show that collective
modes of the ring result in a similar effect of reducing double excitations.
The collective effects also play an important role in Rydberg blockade\cite{Saffman2010Rev.Mod.Phys.2313,Jau2016Nat.Phys.71--74,Labuhn2016Nature}.
The observed anti-bunching in resonance fluorescence in Ref. \cite{Wientjes2014Nat.Commun.}
was originally attributed to the annihilation of two single excitations
. However, we show a rather simple origin of the anti-bunching in
the emitted photon statistics, based on the simple observation from
cyclic structures. In principle, both EEA and photon-blockade could
be responsible for anti-bunching a given experiment. The current mechanism
has not exclude the EEA mechanism, since photon-blockade mechanism
just reduces the rate of double excitation and EEA may take role when
double excitation is accidentally created. 

The current mechanism could be beneficial in light-harvesting processes.
The sunlight energy is mainly absorbed by chlorophylls in these antenna
complexes \cite{VanAmerongen2000}. An excess of energy will produce
an accumulation of excited chlorophylls; which further result in triplet
chlorophylls\cite{DURRANT1990167} and eventually oxidatively damage
chlorophylls\cite{Kaufmann1301,Barber1995,Melis1999}. The blockade
mechanism reduces the probability of two excitations that appears
in the antenna rings, and in turn, separate the output of excitons
in time as sequence. Under this observation, we hypothesise the photon-blockade
as a new mechanism in light-harvesting system to protect the apparatus
from oxidative damage due to excess photons. The blockade mechanism
acts as an additional layer of protection of chlorophylls by reducing
the probability of two excitations in one antenna ring. We remark
that the blockade of photon absorption reduces the probability of
double excitations, however, does not exclude the double excitations. 

\begin{figure}
\includegraphics[width=8cm]{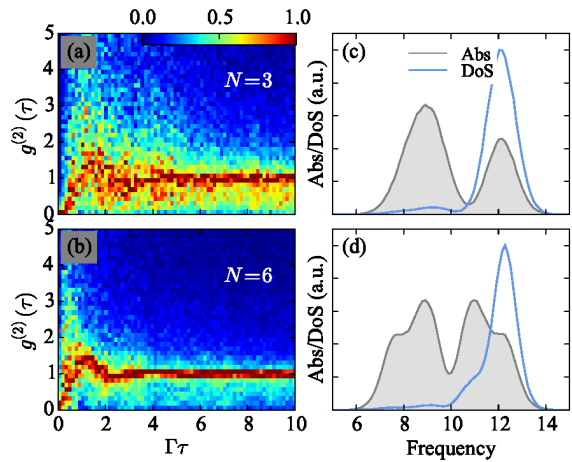}

\caption{(color online) Distribution (a, b) of the second-order correlation
$g^{(2)}(\tau)$ in ensemble system with disorder of both chlorophyll
transition energy $\{\omega_{i}\}$, and the corresponding linear
absorption and density of states for $N=3$ and $N=6$.}

\label{fig3}
\end{figure}

We have illustrated the mechanism of blocking double excitations in
the cyclic structure with very high degree of symmetry ($D_{N}$).
However, the variation of protein environment results in inhomogeneous
broadening (static disorder) and dynamical fluctuation to both chlorophyll
transition energies $\left\{ \omega_{i}\right\} $ and inter-chlorophyll
hopping strength $\left\{ J_{n}\right\} $. It was already proved
that the inhomogeneous broadening of transition energies $\{\omega_{i}\}$
is the main contribution of line-width in absorption spectra of cyclic
structures \cite{Cheng2006Phys.Rev.Lett.028103}. As shown below,
the inhomogeneous broadening simply induces a distribution of second-order
correlation function with the central line matching that of homogeneous
ring. The inclusion of disorder of inter-chlorophyll coupling only
further broadens the distributions. In current study, we will focus
on the transition frequency disorder $\omega_{i}$. To consider static
disorder, we set the transition frequency $\omega_{i}=\omega+\delta\omega_{i}$,
where $\delta\omega_{i}$ is the variation of transition energy on
$n$-th chlorophyll. In our simulation, we choose a Gaussian distribution
for the disorder \cite{Fidder1991J.Chem.Phys.}, namely, 
\begin{equation}
p\left(\delta\omega_{i}\right)=\frac{1}{\sqrt{2\pi}\sigma_{\omega}}\exp[-\frac{\delta\omega_{i}^{2}}{2\sigma_{\omega}^{2}}].
\end{equation}
We examine the second-order correlation of the cyclic structures ($N=3$
and $N=6$), with the disorder \cite{Oijen1999Science} $\sigma_{\omega}=\omega/10$
. The distribution of the second-order correlation is calculated by
sampling of the chlorophyll transition frequencies $\{\omega_{i}\}$
on different chlorophylls on the ring . 

Fig. \ref{fig3}(a) shows the density plot of correlation function
distribution with the chlorophyll-energy disorder on a ring with $N=3$
chlorophylls with inter-chlorophylls dipole-dipole interaction. The
distribution is generated with 1000 sets of energy configuration $\{\omega_{i}\}$.
Even with disorder, the plot for system with dipole-dipole interaction
shows a prominent suppress of two photon emission at $\tau=0$. Fig.
\ref{fig3}(c) shows the corresponding linear absorption spectra (blue
solid line) and the density of state (gray dashed line). The main
peak of the absorption spectrum corresponds to the transition from
the ground state to the superradiance state. The absorption of other
states are suppressed due to the effective weak transition dipole,
despite larger density of state, shown in Fig. \ref{fig3}(c). We
remark that the higher peak on the curve of DoS in Fig. 3(c) is caused
by a two-fold degeneracy of states. The degeneracy is illustrated
in Fig. \ref{fig:setup}(b). 

In Fig. \ref{fig3}(b), we present the correlation function $g^{(2)}(\tau)$
distribution for a ring with $N=6$ chlorophylls. The distribution
shows similar behavior of suppressing simultaneous two-photon emission
as that of the ring with $3$ chlorophylls. Comparing to that with
$N=3$ chlorophylls, the suppressing of double excitation on the ring
is significantly reduced. It was shown that the dense pack of $N$
chlorophylls into unit, smaller than the wavelength it absorbs, increases
the cross section of absorbing photons through similar mechanism of
superradiance. However, we show that the advantage of increasing number
of chlorophylls is trade-off by a increasing the probability of damage,
due to the loss of photon-blockade. Such observation suggests a competing
factor, which could be utilized to understand optimal numbers \cite{Cleary2013Proc.Natl.Acad.Sci.U.S.A.8537--8542}
of pigments in cyclic antennae structures. Another factor is the pure
dephasing rate, due to the dynamical fluctuations of transition energies.
It is known \cite{Clemens2000Phys.Rev.A61_63810} that the anti-bunching
effect is less sensitive to the pure dephasing rate \bibnote{Our numerical calculations for much larger number of atoms $N=8$ and dephasing rate $\gamma_\mathrm{deph}=J/3$,           show that the anti-bunching is still preserved, although less pronounced. (Details avialable on request.)}. 

We have demonstrated the mechanism of photon-blockade with the generic
model of ring structures under a single mode driving field. Replacement
of single mode driving field by sunlight is challenging. This is because
of the broadband nature of sunlight and we need to drive the coefficients
of one- and two-photon absorption processes for the ring with strong
dipole-dipole interaction. 

In conclusion, we have introduced an alternative mechanism - photon-blockade,
which can result in the anti-bunching effect in resonance fluorescence
from natural light-harvesting antennas with cyclic structures, where
each unit has moderate interaction to adjacent ones. This mechanism
can be utilized to explain the anti-bunching behavior observed in
the very recent single-molecule resonant fluorescence experiment on
LHII, where it was speculated as indication of double-excitation annihilation
\cite{Wientjes2014Nat.Commun.}. We further demonstrated the robustness
of the blockade mechanism against the inter-chlorophylls disorder,
which dominantly contributes to the line-width of absorption of light-harvesting
complexes. 
\begin{acknowledgments}
The authors would like to thank Shiyao Zhu, Jianshu Cao, Alexi Sokolov,
Peter Rentzepis, Da-wei Wang, and Da-zhi Xu for helpful discussion.
We gratefully acknowledge support of the National Science Foundation
Grant EEC-0540832 (MIRTHE ERC), the Office of Naval Research, and
the Robert A. Welch Foundation (Award A-1261). GSA thanks the Biophotonics
initiative of the Texas A\&M University for support.
\end{acknowledgments}

\bibliographystyle{apsrev4-1}
\bibliography{SuperRadiance}

\begin{thebibliography}{58}
\bibitem [{1()}]{1}%
  \BibitemOpen
  \href@noop {} {}\bibinfo {note} {Our numerical calculations for much larger
  number of atoms $N=8$ and dephasing rate $\gamma _\mathrm {deph}=J/3$, show
  that the anti-bunching is still preserved, although less pronounced. (Details
  avialable on request.)}\BibitemShut {NoStop}%
\end{thebibliography}

\end{document}